\documentstyle[amsmath,amssymb,epsfig]{mn} 
\title{Drifting sub-pulses in two newly discovered pulsars} 
\author[S. M. Ord, R. Edwards and M. Bailes]
       {S.~M.~Ord,$^1$ 
       R.~Edwards$^1$ and 
       M.~Bailes$^1$\\
$^1$Swinburne University of Technology, \\
        Centre for Astrophysics and Supercomputing \\
        Mail 31 \\
        P. O. Box 218 \\
        VIC 3122 \\
        Australia}

\pagerange{\pageref{firstpage}--\pageref{lastpage}}
\pubyear{2001}

\begin{document} 
\maketitle 

\bibliographystyle{mn}
\begin{abstract}
We have detected the rare phenomenon of stable, drifting sub-pulse behaviour in
two pulsars discovered in the recent Swinburne intermediate latitude pulsar survey. The
pulsars; PSR~J1231--47 and PSR~J1919+0134, have approximate periods ($P$) of 1.873 and
1.6039 seconds respectively.

 Both pulsars have multi-component profiles, and distinct drifting is observed
across them. We have identified a single drift mode in both pulsars:
the drift rate for PSR~J1231-47 being 5.4(1) ms $P^{-1}$ and 5.8(2) ms
$P^{-1}$ for PSR~1919+0134. The drifting is linear across the profile with no
departure from linearity at the edges within the sensitivity of our observations. 

\end{abstract} 
\nokeywords

\section{Introduction}

The Swinburne intermediate latitude survey \cite{ebsb01} has discovered 69
pulsars, 8 of which are recycled. In the process of confirming
candidates a number of pulsars were found to display interesting
emission behaviour, namely pulse nulling and drifting. Two pulsars in
particular, PSR~J1231--47 and PSR~J1919+0134, exhibit
regular drifting sub-pulses.

Drifting sub-pulse behaviour has been considered a litmus test for models of
pulsar emission since the discovery of such periodicities in pulsar emission by
Drake and Craft (1968)\nocite{dc68}. Although many pulsars display sub-pulse
intensity variations few present orderly persistent drifting sub-pulses. It is
considered by Rankin (1986)\nocite{ran86} to be a purely geometrical effect. Assuming the
intensity variations are due to conal sub-beams, then drifting will be apparent
if the pulse profile is a result of an almost tangential cut of the line of
sight across the cone. This implies that drifting sub-pulses would be more
apparent in conal single profiles and that well resolved double profiles should
display intensity variations that do not drift. The newly discovered pulsars are both members
of an intermediate profile class, that of ``barely resolved conal double''
\cite{ran83}. They also display the very rare property of a stable drift pattern.

We present both 2 dimensional auto-correlation analysis (Vivekanand and Joshi
1997\nocite{vj97}) and fluctuation spectra (Backer 1973\nocite{bac73}; Taylor
Manchester and Huguenin 1975\nocite{tmh75}) for both pulsars; providing an
initial characterisation of their sub-pulse behaviour. We also present an
examination of the drift rate of an average drift band as a function of pulse
phase.

\section{Observations}

All the observations presented here were taken with the Parkes 64 metre radio
telescope using the central beam of the 13 beam multi-beam receiver and the
$96\times3$ MHz filter bank (Staveley--Smith et al 1996\nocite{swb+96}), at a central
frequency of 1374 MHz. Both polarisations were summed together and the
result one bit sampled.  The sample rate varies between the  observations,
and is 125 $\mu s$ for PSR~J1919+0134 and  500 $\mu s$ for PSR~J1231-47. This
time series is folded into 512 phase bins at the topocentric pulsar spin period.

\begin{figure*}
\psfig{file = 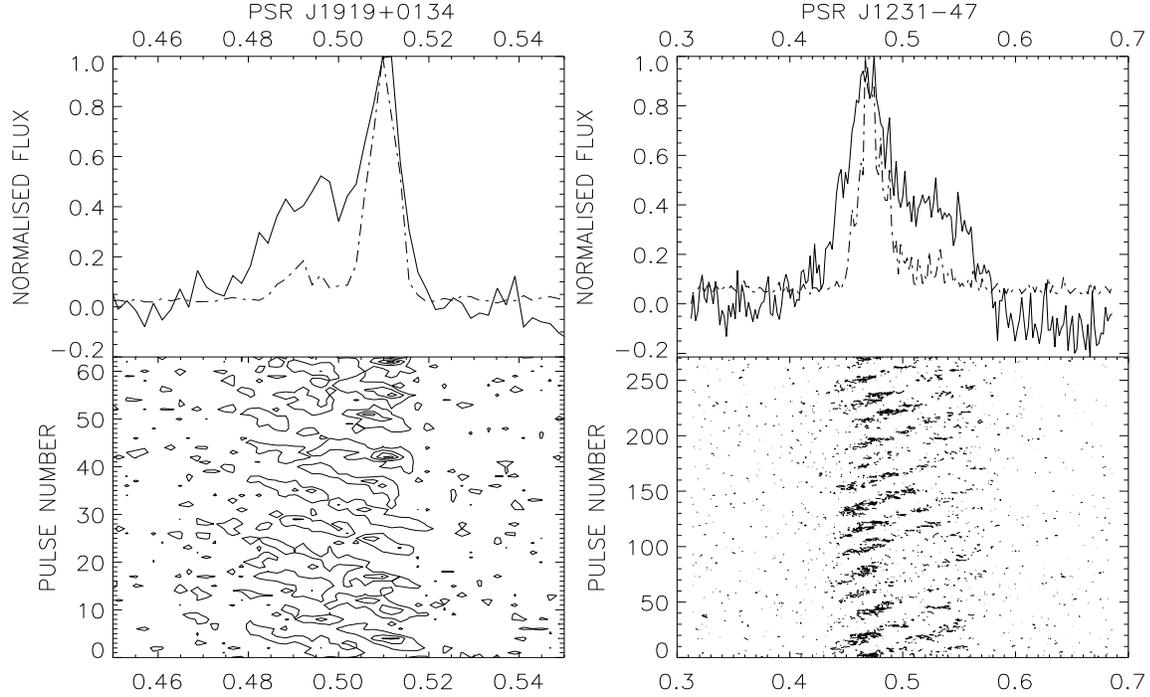,width = 15cm,height=10cm}
\caption{Longitude--time plot showing the drifting sub-pulses of J1231+47 and
J1919+0134. The contour levels have been set to a scale that increases the
contrast.  Phase 0 has been set arbitrarily. The dotted line indicates the
level of spectral power in the fluctuation spectrum at that longitude.}
\label{2panel} 
\end{figure*}

Each period can be represented by a row in a ``longitude--time'' plot
\cite{tmh75,mt77}. Two such plots are shown in Figure \ref{2panel}
together with average profiles for both pulsars. These arrays form the basis of
the following analysis. 

Many different methods have been employed to examine drifting behaviour, including cross--correlation \cite{pw86}, Fourier phase methods and fluctuation spectra
(Backer 1973, Deshpande and Rankin 2001\nocite{bac73}\nocite{dr01}) and
auto-correlation methods \cite{vj97}.  Both fluctuation spectra and
auto-correlation analyses are presented here.  

\begin{table*} 

\caption{Astrometric, spin and dispersion parameters for the pulsars under
investigation, from Edwards et. al. (2001). It should be noted that there is no complete timing solution for PSR~J1231--47. }

\label{positions} 
\begin{tabular}{||l|l|l|l|l|l|l||} 
\hline 
Name & R.A. (J2000) & Dec. (J2000) & $P$ & $P$ Epoch & $\dot{P}$ & DM \\ 
& & & (s) & (MJD) & ($10^{-15}$) & (cm$^{-3}$ pc)\\

J1231--47 & 12$^{\mathrm h}$31$^{\mathrm m}$40(50) &
--47\degr46(7)\arcmin & $1.87304$(2) & 52056.3 & \dotfill & $31$(3)\\

J1919+0134 & 19$^{\mathrm h}$19$^{\mathrm m}$43\fs62(3) &
+01\degr34\arcmin56\farcs5(7) & $1.60398355528$(6) & 51650.0 &
$0.589$(11) & $191.9$(4)\\

\hline

\end{tabular} 

\medskip

\end{table*}
\subsection{Fluctuation spectra}

Fluctuation spectra were formed by performing a spectral
analysis of the longitude--time data. The process is discussed in
 Backer \shortcite{bac73}. The Fourier transform was performed upon each
longitudinal column to produce an indication of the level of
periodicity as a function of pulse phase. The spectra were then 
normalised by first subtracting the power spectra of an off-pulse
region from all the on-pulse spectra. Each spectrum was then divided by
the square of the average signal level in the longitudinal column from
which the spectrum was formed.  Figure \ref{2panel} shows  the
normalised average pulse profile along with the normalised spectral power as
a function of longitude measured at the peak of the fluctuation
spectrum.  Both pulsars are in fact very similar. The fluctuation, although
great in longitudinal extent, is concentrated in the stronger of the two
components.

 Analysis of the spectra themselves 
provided values for the frequency of the amplitude variations. The values
obtained for the frequency of the main spectral feature were: $0.050 \pm 0.002$
(cycles/period) for PSR~J1231--47, and $0.156 \pm 0.008$ (cycles/period), for
PSR~J1919+0134. 

In order to further investigate this phenomenon, a 1 hour integration of
PSR~J1231-47 has been obtained. Due to interstellar scintillation effects the
1 hour observation had a signal to noise ratio such that sub-pulses are not
directly observable in a longitude--time plot. Furthermore it suffered from a
burst of interference approximately half way through. In order to mitigate the
interference effects the observation was broken into two sub--integrations.
Nevertheless fluctuation spectra formed from these sub--integrations displayed a
significant spectral feature at $0.0528 \pm 0.0008$ (cycles/period) and $0.0515
\pm 0.0004$ (cycles/period) respectively. This feature is consistent with that
found in the shorter observation.  The quoted errors are under--estimates being
simply half a frequency bin-width.  These values are consistent with those
presented in Table \ref{properties}

\subsection{Average properties}
\label{props}

In order to examine the features of a ``typical'' drift band some sort of band
averaging had to be performed.  This was achieved by the construction of  
a two dimensional auto-correlation function (2DACF) of the longitude -- time plots
\cite{vj97}.  The auto-correlation function was formed in a number of stages. A
region of the longitude time frame was chosen. The region is composed of the
on--pulse region in longitude domain and the length of the observation in the
time domain. The mean value of this sub--array was then subtracted from each
point. The purpose of this was to reduce the height of the zero--lag spike in
the auto-correlation. The array was then padded on all 4 sides with a number of
zeroes commensurate with the number of lags required in the 2DACF. A forward
Fourier transform was then performed. The resultant was
then multiplied by its complex conjugate. The product was then inverse
transformed, producing the 2DACF.

The 2DACF provides a method for examining the ``average'' properties of the
drifting bands by providing the correlation of each drift band with itself and its
neighbours. We can examine the evolution of the drift rate by comparing the
slopes and separations of the correlated bands. The values for $P_{2}$ (the
phase separation), and $P_{3}$ (period separation) \cite{bac73} can be found by
analysis of straight line fits to the peaks in the lag-lag plane.  The straight line fits
were performed on the central regions of each drift band, where the signal to noise 
ratio was highest. In order to obtain the parameters of the straight line fit
we assumed the data was adequately modeled by a straight line and an initial
fit was performed.  Then measurement errors of magnitude 1 standard deviation
were ascribed to each point, the fit performed again and the chi-square
statistic minimised to  obtain values and errors for the fitted parameters.
This method is described in Press et al. (1986)\nocite{pftv86}.

The values for the separation between pairs of bands in time ($P_{3}$) and
phase ($P_2$) are given in Table \ref{properties}. The values were obtained by
combining the values for individual drift bands in the lag--lag plane obtained
from the straight line fits.

\subsection{Drift rate across the profiles} \label{drift}

Many pulsars which display the property of drifting sub-pulses exhibit a change
in sub-pulse drift rate across the pulse profile. This property may be used to
test emission models \cite{kri80,os76,wri81}.

We have examined the straight line fits to the peaks in the lag--lag plane in
an attempt to identify any clear departure from linearity. No discernible,
regular structure indicative of a changing drift rate is apparent in the
residuals. Although it should be noted that the ``grand averaging'' properties of
the 2DACF method will remove any non-linearity if the drift rate varies between
bands. The low signal to noise ratio of the sub-pulses at the edges
of the drift band make investigation of the band behaviour in this region difficult.

\begin{table} 
\caption{ The parameters of a ``representative'' or ``average'' drift
band for the two pulsars. The parameters were derived from analysis of the straight line
fits to peaks in the 2 dimensional auto-correlation functions.
$P_2$ is the separation of two drift bands in phase. $P_3$ is the separation in 
periods of the drift bands. The drift rate is given as the number of periods 
required for an average drift band to cross unit phase. The drift rate in milliseconds per period is also given.
 }
\label{properties} 
\begin{tabular}{||l|l|l|l|l||} 
\hline 
Name & $P_2$ ($\phi$) & $P_3$ ($P$) & Drift rate ($P/\phi$) & ($ms/P$)\\

J1231--47 & 0.0557(9) & 19.3(2) & 348(8) & 5.4(1) \\
J1919+0134 & 0.023(1) & 6.5(2) & 277(8) & 5.8(2) \\

\hline

\end{tabular} 

\end{table}

\section{Discussion}

Very regular, stable, persistent drifting pulsars are rare. As such these
pulsars represent a welcome addition to the pulsar family. The initial
observations are intriguing, the linearity of the drift may provide some
information regarding emission geometry \cite{kri80,wri81}. The longitudinal extent
of the drift is interesting; it is rare for
pulsars of this profile class to display a measurable drift across the whole
profile \cite{ran86,hw87}. Both of these pulsars display a measurable $P_3$ across the majority
of the profile and the drift is continuous. Furthermore it is rare for pulsars
of this profile class to show such a clear sense of drift, as simple periodic
intensity variations at fixed longitudinal positions are more common.  

It appears that these pulsars are rare and exemplary exponents of the drifting
sub pulse phenomena; as such these pulsars will be of great benefit in
investigations of the radio emission mechanism in pulsars.

\bibliography{mnrasjournals,modrefs,psrrefs,crossrefs} 

\begin{thebibliography}{{Taylor, Manchester \& Huguenin }{1975}}

\bibitem[\protect\citename{Backer }{1973}]{bac73}
Backer~D.~C., 1973, Astrophys.\,J., 182, 245

\bibitem[\protect\citename{{Deshpande} \& {Rankin} }{2001}]{dr01}
{Deshpande}~A.~A., {Rankin}~J.~M., 2001, Mon.\,Not.\,R.\,astr.\,Soc., 322, 438+

\bibitem[\protect\citename{Drake \& Craft }{1968}]{dc68}
Drake~F.~D., Craft~H.~D., 1968, Nature, 220, 231

\bibitem[\protect\citename{{Edwards} {\rm et~al. }}{2001}]{ebsb01}
{Edwards}~R.~T., {Bailes}~M., {van Straten}~W., {Britton}~M.~C., 2001,
  Mon.\,Not.\,R.\,astr.\,Soc., 326, 358+

\bibitem[\protect\citename{Hankins \& Wolszczan }{1987}]{hw87}
Hankins~T.~H., Wolszczan~A., 1987, Astrophys.\,J., 318, 410

\bibitem[\protect\citename{Krishnamohan }{1980}]{kri80}
Krishnamohan~S., 1980, Mon.\,Not.\,R.\,astr.\,Soc., 191, 237

\bibitem[\protect\citename{Manchester \& Taylor }{1977}]{mt77}
Manchester~R.~N., Taylor~J.~H., 1977, Pulsars.
\newblock Freeman, San Francisco

\bibitem[\protect\citename{Oster \& Sieber }{1976}]{os76}
Oster~L., Sieber~W., 1976, Astrophys.\,J., 210, 220

\bibitem[\protect\citename{Press {\rm et~al. }}{1986}]{pftv86}
Press~W.~H., Flannery~B.~P., Teukolsky~S.~A., Vetterling~W.~T., 1986, Numerical
  Recipes: {T}he Art of Scientific Computing.
\newblock Cambridge University Press, Cambridge

\bibitem[\protect\citename{Proszynski \& Wolszczan }{1986}]{pw86}
Proszynski~M., Wolszczan~A., 1986, Astrophys.\,J., 307, 540

\bibitem[\protect\citename{Rankin }{1983}]{ran83}
Rankin~J.~M., 1983, Astrophys.\,J., 274, 333

\bibitem[\protect\citename{Rankin }{1986}]{ran86}
Rankin~J.~M., 1986, Astrophys.\,J., 301, 901

\bibitem[\protect\citename{Staveley-Smith {\rm et~al. }}{1996}]{swb+96}
Staveley-Smith~L. {\rm et~al.}, 1996, Proc.\,Astr.\,Soc.\,Aust., 13, 243

\bibitem[\protect\citename{Taylor, Manchester \& Huguenin }{1975}]{tmh75}
Taylor~J.~H., Manchester~R.~N., Huguenin~G.~R., 1975, Astrophys.\,J., 195, 513

\bibitem[\protect\citename{Vivekanand \& Joshi }{1997}]{vj97}
Vivekanand~M., Joshi~B.~C., 1997, Astrophys.\,J., 477, 431

\bibitem[\protect\citename{Wright }{1981}]{wri81}
Wright~G. A.~E., 1981, Mon.\,Not.\,R.\,astr.\,Soc., 196, 153+

\end{thebibliography}

\begin{figure*} 
\psfig{file = 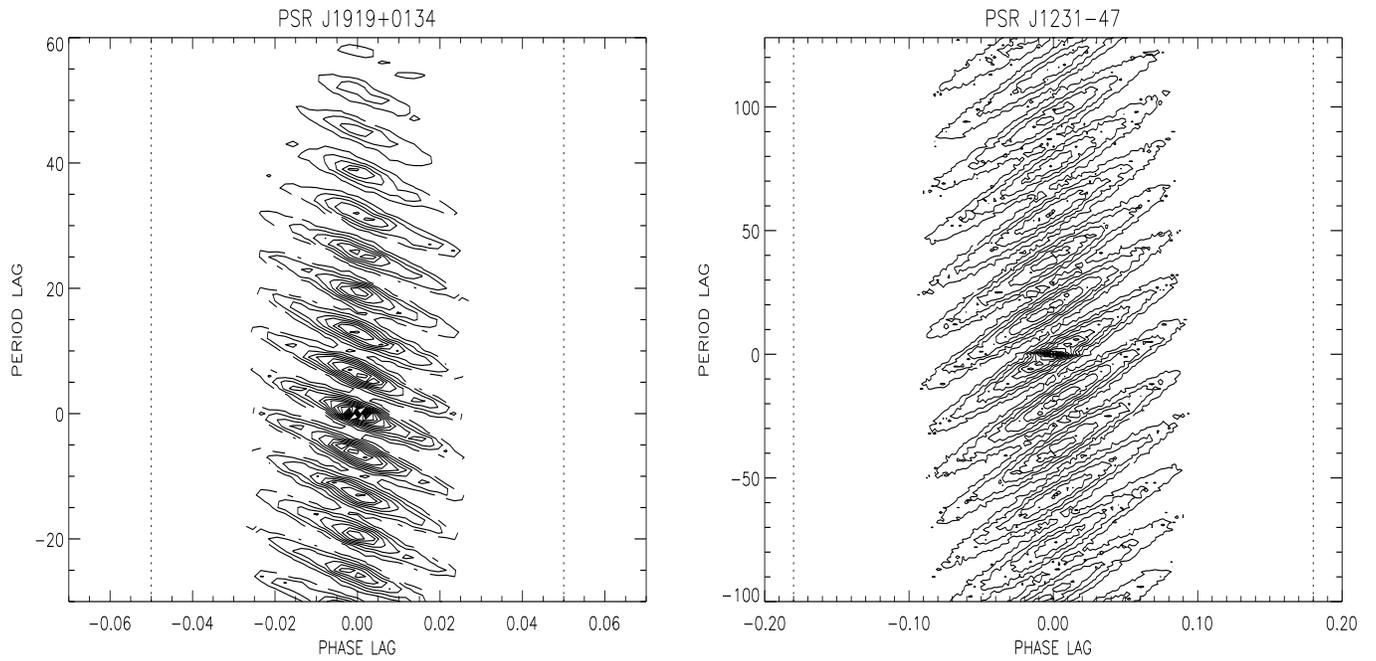,width = 20cm,height=9cm}
\caption{The 2 dimensional autocorrelations of the longitude--time information.
A straight line has been fit to the central peaks in the lag-lag plane in order to 
characterise the drift band behaviour. The parameters are given in Table \ref{properties}. 
} \label{2DACF}
\end{figure*}

\end{document}